\documentclass[conference]{IEEEtran}
\IEEEoverridecommandlockouts
\usepackage{cite}
\usepackage{amsmath,amssymb,amsfonts}
\usepackage{algorithmic}
\usepackage{graphicx}
\usepackage{hyperref}
\usepackage{textcomp}
\usepackage{xcolor}
\def\BibTeX{{\rm B\kern-.05em{\sc i\kern-.025em b}\kern-.08em
    T\kern-.1667em\lower.7ex\hbox{E}\kern-.125emX}}
\begin{document}

\newif\ifdraft
\drafttrue
\ifdraft
\definecolor{ocolor}{rgb}{1,0,0.4}
\newcommand{\jhanote}[1]{ {\textcolor{red} { ***shantenu: #1 }}}
\else
\newcommand{\jhanote}[1]{}
\fi

\title{DeepDriveMD: Deep-Learning Driven Adaptive Molecular Simulations for
Protein Folding}


\author{
Hyungro Lee$^{1*}$, Heng Ma$^{2*}$, Matteo Turilli$^{1*}$, Debsindhu Bhowmik$^{3}$, \\ 
Shantenu Jha$^{1,4**}$, Arvind Ramanathan$^{2**}$\\
   \footnotesize{\emph{$^{1}$RADICAL, ECE, Rutgers University, Piscataway,NJ 08854, USA}}\\
   \footnotesize{\emph{$^{2}$Argonne National Laboratory}} \\
   \footnotesize{\emph{$^{3}$Oak Ridge National Laboratory}} \\
   \footnotesize{\emph{$^{4}$Brookhaven National Laboratory, Upton, NY, USA}}\\
   \footnotesize{\emph{$^{*}$Joint First Authors}} \\
   \footnotesize{\emph{$^{**}$Senior Authors}}
  }



\maketitle

\begin{abstract}
Simulations of biological macromolecules play an important role in
understanding the physical basis of a number of complex processes such as
protein folding. Even with increasing computational power and evolution of
specialized architectures, the ability to simulate protein folding at
atomistic scales still remains challenging. This stems from the dual aspects
of high dimensionality of protein conformational landscapes, and the inability
of atomistic molecular dynamics (MD) simulations to sufficiently sample these
landscapes to observe folding events. Machine learning/deep learning (ML/DL)
techniques, when combined with atomistic MD  simulations offer the opportunity
to potentially overcome these limitations by: (1) effectively reducing the
dimensionality of MD simulations to automatically build latent representations
that correspond to biophysically relevant reaction coordinates (RCs), and (2)
driving MD simulations to automatically sample potentially novel
conformational states based on these RCs. We examine how coupling DL
approaches with MD simulations can fold small proteins effectively on
supercomputers. In particular, we study the computational costs and
effectiveness of scaling DL-coupled MD workflows by folding two prototypical
systems, viz., Fs-peptide and the fast-folding variant of the villin head
piece protein. We demonstrate that a DL driven MD workflow is able to
effectively learn latent representations and drive adaptive simulations.
Compared to traditional MD-based approaches, our approach achieves an
effective performance gain in sampling the folded states by at least 2.3x. Our
study provides a quantitative basis to understand how DL driven MD
simulations, can lead to effective performance gains and reduced times to
solution on supercomputing resources.


\end{abstract}


\section{Introduction}


Understanding the biophysical processes that control how a polypeptide folds into its three-dimensional native structure remains an outstanding question in molecular biology. Experimental studies,  simulations and theory have continued to provide valuable insights into how proteins fold, especially in the context of small, and fast folding proteins (typical folding times of about several $\mu$s-ms).~\cite{adcock2006molecular} It is generally accepted that proteins fold through a discrete number of intermediate states, where each state consists of partial folded components in terms of secondary structures~\cite{Englander_2014}. Associated with these intermediate states are timescales that characterize their stability (either in terms of how long a secondary structure may persist or other physical/ structural attributes) are before the protein `jumps' into other states finally reaching its folded state.

The inherent high dimensionality of protein folding trajectories (generated
from simulations) makes it challenging to characterize: (i) metastable states
-- states  that share similarity in structure/ conformation, and other
biophysical properties of interest, and (ii) transition times -- how stable
these intermediate states. A number of clustering approaches have therefore
been developed for obtaining insights into metastable states and
characterizing transition
times~\cite{Shao_2007,Keller_2010,Yan_2016,Shukla_2015}. Such approaches build
reduced dimensional (latent) representations from molecular dynamics (MD)
data, typically using principal component analysis or independent component
analysis techniques~\cite{scherer_pyemma_2015,Ramanathan_2011}.

Complementary to such approaches, we recently developed a deep convolution
variational autoencoder (CVAE)~\cite{bhowmik2018deep}, to automatically
cluster protein folding trajectories into a small number of conformational
states. Our approach was able to organize the conformational landscape based
on key reaction coordinates for protein folding such as the fraction of native
contacts and the root mean squared deviations (RMSD) to the native state.
Further, our approach also allowed us to transfer these learned properties
across independent simulations.

In addition to the aforementioned challenges, MD simulations tend to get
`stuck' within metastable states~\cite{Shukla_2015}; a variety of approaches
have been developed to address this challenge. These techniques, collectively
referred to as enhanced sampling methods, use: (1) a pre-determined set of low
dimensional representations referred to as reaction coordinates or collective
variables determined from MD simulations either biophysically determined {\it
a priori}~\cite{grubmuller1996ligand, abrams2010large, kastner2011umbrella}
(e.g., distances between key residues~\cite{kastner2011umbrella}), or by
learning latent dimensional representations (e.g., described above) to
adaptively sample and accelerate protein folding/ or other biophysical
phenomena of interest, and/or (2) importance sampling techniques that enhance
`rare' events in the simulations.

\begin{figure}
    \centering
    \includegraphics[width=0.5\textwidth]{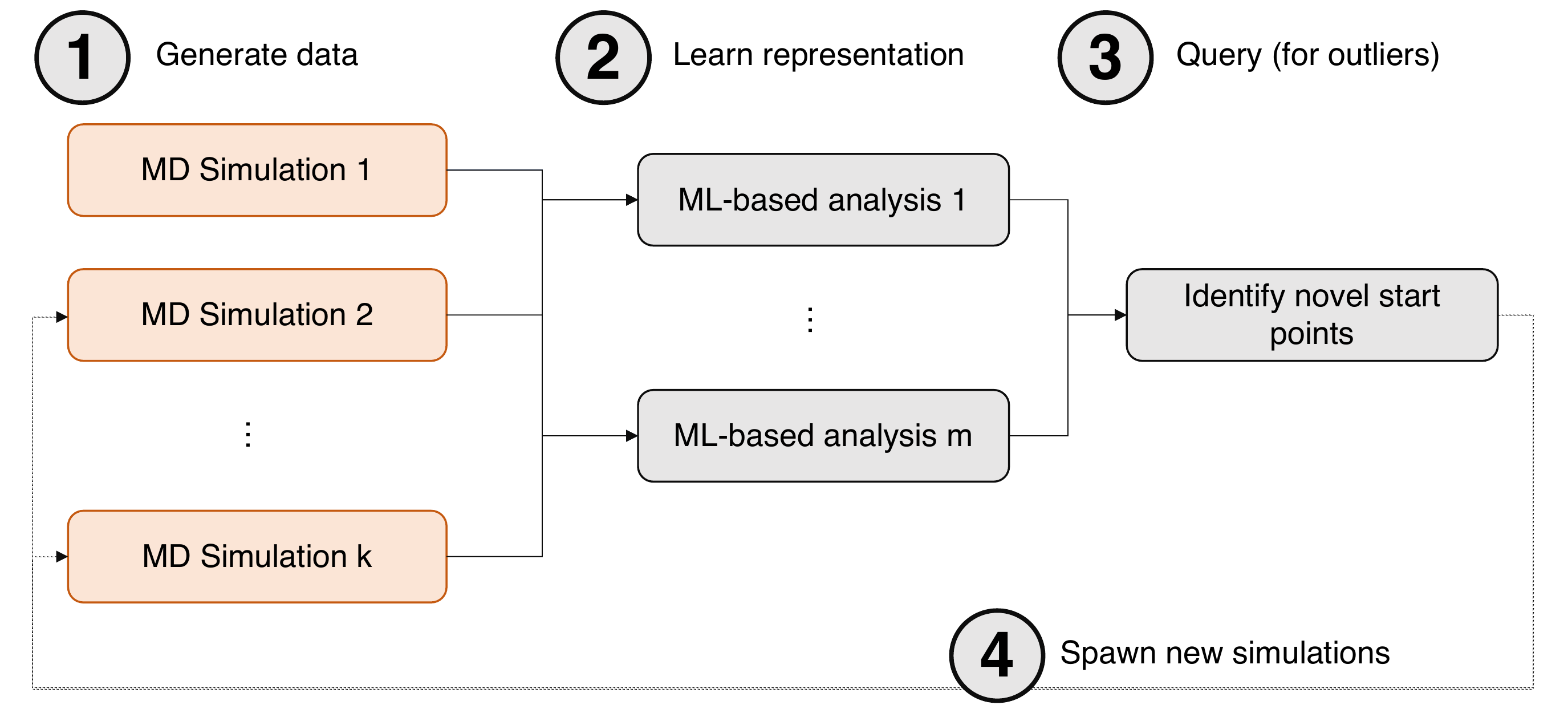}
    \caption{Computational workflow `motif' for coupling MD simulations with ML approaches.}
    \label{fig:workflow-meta}
\end{figure}

Most enhanced sampling techniques~\cite{reap2018shukla, huber1996weighted}
involve generating an initial pool of MD data (either a single simulation or
an ensemble of simulations), followed by intermittently stopping and steering
MD simulations towards novel starting points~\cite{Kasson2018-sw}. We recently
generalized the above workflow~\cite{ma2019deep}  to include learning driven
MD simulations, which we now formalize into a computational motif (see Fig.
\ref{fig:workflow-meta}) and implement at scale. The motif is characterized
by: (1) generating an initial pool of MD data typically using a large ensemble
of MD simulations, followed by (2) a `training' run consisting of a ML
algorithm, and (3) an `inference' step where novel starting points for MD are
identified and (4) new MD simulations are spawned. This may include either
starting entirely new simulations (i.e., expanding the pool of initial MD
simulations) or killing unproductive MD simulations (i.e., simulations that
seem to be stuck in metastable states). We note that this computational motif
represents many scientific discovery processes beyond protein folding.

In general, there are three primary motivations for the coupling of ML/DL
driven approaches with traditional HPC simulations: (i) ML/DL can be used to
reduce the computational cost of simulating a process via the creation of a
computational surrogates; (ii) improve the effective performance of vanilla
HPC simulations by using ML/DL driven HPC simulations, and (iii) use
simulations to improve the training of ML/DL models (both in the presence of
sparse data and otherwise). In this paper, we investigate the latter two
scenarios and discuss software systems solution that provide generalized
implementation and capabilities.

There are at least three different measures of performance of DL driven MD
simulations of relevance: (A) Traditional measures of scale, scalability, and
efficiency; (B) performance of a DL driven simulation method (or algorithm)
relative to a  non-DL driven ``pure'' simulation based method; and (C) the
performance of the learning component as a function of the number of
simulation elements coupled to the learning component. This paper discusses
all three performance measures. However, it is difficult to assess performance
measure C {\it a priori}, which in turn, influences the optimal partition
between resources assigned to the ML/DL component and those assigned to HPC
simulations. Performance measure C also influences performance measure B, and
possibly also efficiency and scalability. Our software system has the ability
to dynamically partition and balance resources assigned to the ML component
and those assigned to HPC component to optimize performance measure C.




The contributions of this paper are three fold: (1) We design and implement
DeepDriveMD --- a framework for deep learning driven simulations using
RADICAL-Cybertools. DeepDriveMD is not constrained to specific learning
methods (e.g., CVAE) but can support arbitrary deep learning driven methods
and HPC simulations;  (2) We utilize DeepDriveMD for a VAE driven adaptive
molecular simulations and show that it is possible to fold small
proteins/peptides using on Summit --- a leadership computing platforms at Oak
Ridge National Laboratory; and (3) We provide performance measures for
integrated learning-simulation methods, assessing the overall effectiveness of
our workflow relative to non-DL driven simulations.




\section{Related Work}

Notable examples of implementation of the workflow motifs --- partial and
complete, include the REAP approach~\cite{reap2018shukla}, where the authors
define a mapping between a finite set of states and actions that enable an
agent to achieve its goal based on a user defined set of order parameters
(OPs) /reaction coordinates (RC). The weights on the RCs is initialized, with
MD simulations used to learn which RCs contribute most to the final target.
Similarly, Galvelis and Sugita defined an enhanced sampling
protocol~\cite{galvelis2017neural} that is based on nearest neighbor density
estimator and a neural network to define a bias potential that resulted in
ergodic sampling and characterizing free energy profiles for various
polypeptides. Notably, this approach currently seems to be limited to
8-dimensional bias potentials. Wang, Ribeiro, and Tiwary use a
VAE~\cite{ribeiro2018reweighted} similar to our approach, however, constrain
the encoder/decoder with an information bottleneck that identifies an optimal
RC. Other approaches such as the neural networks-based variationally enhanced
sampling~\cite{Bonati_2019} and Boltzmann Generators~\cite{noe2019boltzmann}
share similar workflow motifs, although the exact use of MD simulations versus
other types of sampling (e.g., Markov chain Monte Carlo) may differ.

The approach investigated in this paper, and the other approaches described
above are different from AlphaFold~\cite{Senior_2018}, where the target
problem is to model the final folded 3-dimensional structure of a protein from
its primary sequence.

The distinction between this work and aforementioned implementations of the
motif are: (i) Methodological enhancement: this work investigates the
interplay between simulations and learning. Specifically, it can adaptively
tune the ratio of computational resources assigned to simulations and learning
based upon Reconstruction loss;  (ii) Scale: thanks to first-order middleware
for HPC workflows, the scale of problem investigated is significantly greater
than previously reported, and (iii) Generality: Our proposed motif and
software system can support multiple learning methods. We demonstrate the
impact using CVAE based DL method, but could just as well use other learning
methods.

\section{Computational Problem, Design and Implementation}
The overall scientific goal of our paper can be summarized as follows: given an initial set of starting conformations, representing the unfolded state ensemble of a protein, run simulations to enable an efficient sampling of the final folded state using MD simulations. 
In the workflow, MD simulations are overseen by the CVAE model that collects the MD conformers as training input and in return identifies the state of each simulation for interative decision-making on whether to continue or terminate an individual MD task. Conformers in less populated latent space of the CVAE representation are selected as `outliers' for instantiating a new MD task. The outliers are inferred on the basis of using the density based spatial clustering of applications with noise (DBSCAN) algorithm in the latent dimensions of the CVAE model with the lowest reconstruction loss~\cite{bhowmik2018deep}. Note that there are several choices for the selection of outliers; we used one that is known to work well in practice. The number of outliers identified is capped at 150 with a maximum of 10 members in each cluster (identified by DBCSAN). This is reasonable on the basis of the number of initial simulations carried out. 

The workflow also requires setting up the MD and ML tasks, managed by a contemporary scheduler to administer the computational resource and enable interfacing between MD simulations and DL framework, on the specific architecture of Summit, an IBM AC922 system that integrates more than 27,000 NVIDIA V100 GPUs and
9,000 IBM Power9 CPUs. Note that on Summit, each node consists of 2 CPUs with 6 fully inter-connected GPUs using the NVLINK architecture. 
Despite the computational demanded fulfilled with Summit HPC system, the workflow also requires setting up the MD and ML tasks, managed by a contemporary scheduler to administer the computational resource and enable interfacing between MD simulations and DL framework on Summit. 
The MD task is carried out by GPU-accelerated OpenMM molecular simulation engine and VAE framework is set up with Keras/TensorFlow, also on GPU. 
Both tasks enable the workflow to fully leverage the GPU nodes on Summit. 
Here we adopt the RADICAL-Cybertools to conduct all the tasks of the workflow in a scalable fashion. 

\subsection{RADICAL-Cybertools: Ensemble Execution on Summit}
\label{ssec:cybertools}

The RADICAL-Cybertools (RCT) software stack is used to support the scalable concurrent and sequential execution of heterogeneous tasks on high-performance computing (HPC) resources. RCT are a set of software systems that serve as middleware to develop efficient and effective tools for scientific computing. Specifically, RCT enable executing ensemble-based applications at extreme scale~\cite{turilli2019characterizing} and on a variety of computing infrastructures. 


RCT consists of three main components: RADICAL-Ensemble Toolkit (EnTK)~\cite{entk-icpp-2016,balasubramanian2018harnessing}, RADICAL-Pilot (RP)~\cite{merzky2018using}, and RADICAL-SAGA(RS)~\cite{saga-x,ogf-gfd-90}. EnTK provides the ability to create and execute ensemble-based workflows/applications with diverse coordination and communication algorithms, abstracting the need for explicit resource management. EnTK uses RP as a pilot-based~\cite{turilli2017comprehensive} runtime system to provide resource management and task execution capabilities. In turn, RP uses RS as an access layer towards HPC resources.

RCT adopts the ``building blocks'' approach to workflows~\cite{turilli2019middleware,balasubramanian2019radical,jha2019incorporating}. RCT provide scalable implementations of building blocks in Python and are currently used to support dozens of scientific projects on HPC systems, including several existing and prior INCITE awards. RCT is increasingly being used to support applications that involve the concurrent and adaptive execution of ML and simulation tasks~\cite{fox2019understanding}. RCT has been used extensively to support biomolecular sciences algorithms/methods, e.g., replica-exchange, adaptive sampling and high-throughput binding affinity calculations.

\begin{figure}[ht!]
  \centering
  \includegraphics[trim=0 0 0 0,clip,width=0.49\textwidth]{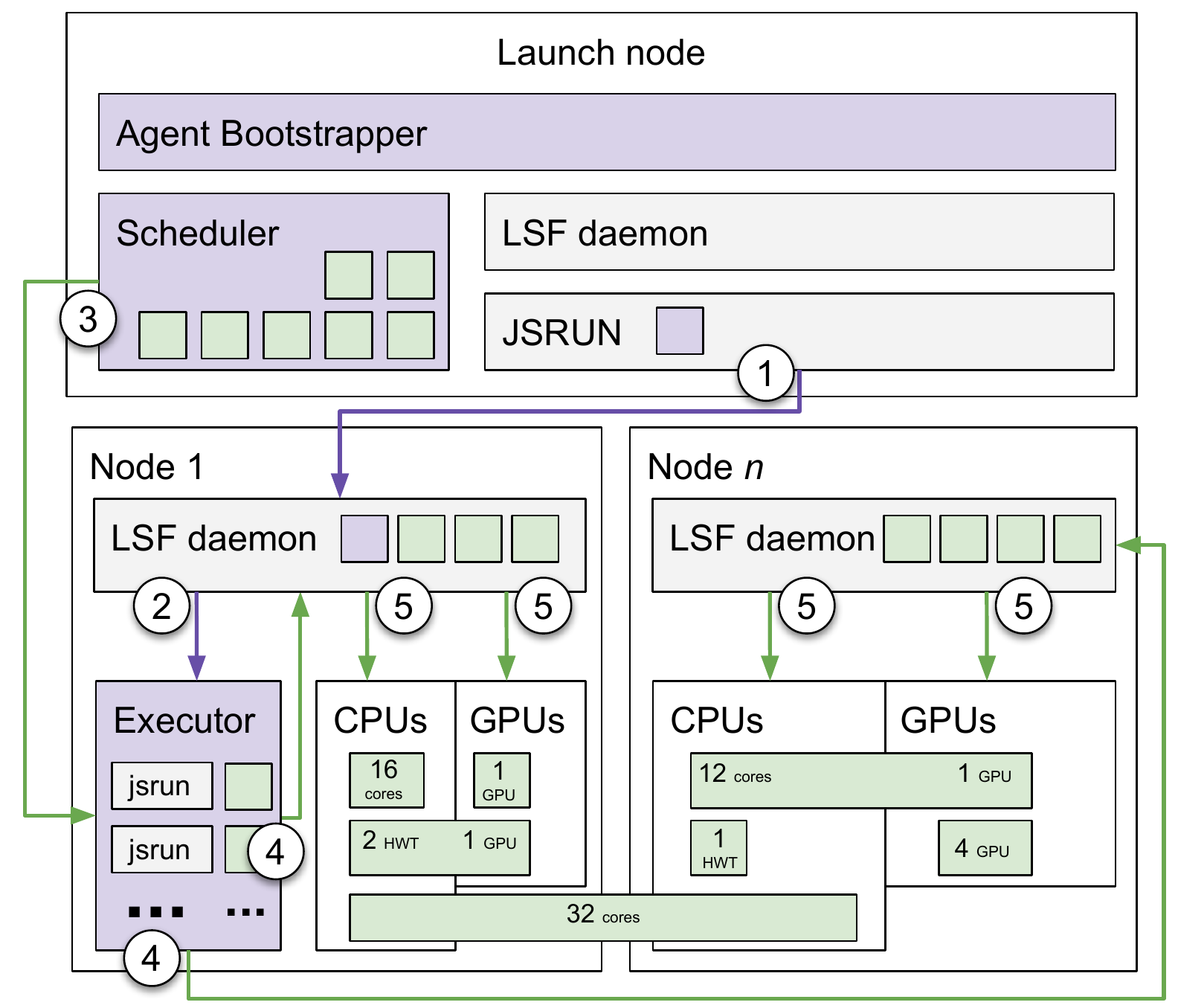}
  \caption{RADICAL-Pilot (RP) deployment on Summit. Purple: RP components; Gray: IBM Platform Load Sharing Facility (LSF) components; Green: heterogeneous computational tasks. 1-2: scheduling of RP's Executor component on a work node via LSF daemons and JSRUN; 3-5: scheduling of computational tasks via RP's Scheduler component, LSF daemons and JSRUN. RP manages heterogeneous tasks that require arbitrary combinations of available resources.}
\label{fig:rp-summit}
\end{figure}



RP is an implementation of the pilot abstraction, engineered to support scalable launching of heterogeneous tasks across different HPC platforms. RP is a runtime system designed to decouple resource acquisition from task execution. As every pilot system, RP acquires resources by submitting a batch job, then bootstraps dedicated software components on those resources to schedule, place and launch application tasks, independent from the machine batch system.


RP is a distributed system designed to instantiate its components across available resources, depending on the platform specifics. Each components can be individually configured so as to enable further tailoring while minimizing code refactoring. RP uses RS to support all the major batch systems, including Slurm, PBSPro, Torque and LSF\@. RP also supports many methods to perform node and core/GPU placement, process pinning and task launching like, for example, aprun, JSM, PRRTE, mpirun, mpiexec and ssh.

RP is composed of two main components: Client and Agent. Client executes on any machine while Agent bootstraps on one of Summit's batch nodes. Agent is launched by a batch job submitted to a batch system via RS. After bootstrapping, Agent pulls bundles of tasks from Client, manages the tasks' data dependencies if any, and then schedules tasks for execution via one or more launching methods. RP can execute scalar, OpenMP, MPI tasks within and across multiple nodes, allowing each task to use one or more CPU/GPU exclusively or concurrently. 

RP has been ported to Summit enabling fine-grained mapping, scheduling and execution of heterogeneous computational tasks on CPU, GPU, and hardware threads (HWT). 
Agent deployment depends on several configurable parameters like, for example, number of sub-agents, number of schedulers and executors per sub-agent, and method of placing and launching tasks for each executor of each sub-agent. On Summit, the default deployment of Agent instantiates a single sub-agent, scheduler and executor on a batch node. The executor calls one \texttt{jsrun} command for each task, and each \texttt{jsrun} uses the JSMD demon to place and launch the task on work nodes resources (thread, core and GPU).

Fig.~\ref{fig:rp-summit} shows an alternative deployment of Agent that uses PRRTE/DVM instead of JSM/LSF. 
to place and launch tasks across compute nodes. This configuration enables a sub-agent to use more resources than with JSM/LSF and improves scalability and performance of task execution. Note that, independent from the configuration and methods used, RP can concurrently place and launch different types of tasks that use different amount and types of resources. Our tests show reliable concurrent execution of up to 16384 tasks, each task using 1 core for a total of 404 compute nodes, and up to 100 tasks, each requiring 1096 cores.



EnTK exposes an application programming interface (API) for the description of scientific applications as static or dynamic sets or sequences of pipelines. Each pipeline is composed of stages and each stage contains an arbitrary set of tasks. Tasks can execute concurrently while stages can execute only sequentially. These properties are insured by design, offering what we have called a Pipeline Stage Task (PST) model for the specification of computational workflows. It is important to note that 'task' here are not functions, methods or sub-processes of one of EnTK components. Task indicates instead a self-contained process (i.e., program) executed and managed by the operating system of the target resource. Consistently, tasks can be a single-threaded, multi-threaded or MPI program, and can use CPUs, GPUs or both within and across the compute nodes of a target machine.

\subsection{Integration of ML and MD}

Many scientific workloads are comprised of many tasks, where each task is an independent simulation or data processing analysis. The execution of many tasks on heterogeneous HPC platforms requires scalable dynamic resource management and multi-level scheduling. Together, EnTK and RP enable the codification of many-task applications and their scalable execution on HPC machines like Summit.

In a recent paper~\cite{turilli2019characterizing}, we characterized the performance of executing many tasks using RP when interfaced with JSM or PRRTE on Summit: RP is responsible for resource management and task scheduling on acquired resource; JSM or PRRTE enact the placement and launching of scheduled tasks. When using homogeneous single-core, 15 minutes-long tasks, PRRTE scales better than JSM for $>$ O(1000) tasks; PRRTE overheads are negligible; and PRRTE supports optimizations that lower the impact of overheads and enable resource utilization of 63\% when executing O(16K) 1 core tasks over 404 compute nodes. In this paper, the workload is comprised of heterogeneous tasks of varying temporal durations but the resource utilization and scaling remain invariant.

For each experiment of this paper, we vary only the number of starting conformations, i.e., how many simulations are initiated across multiple GPUs on Summit. We explicitly choose only one Summit node, training our CVAE model on 4 out of the 6 GPUs available. This is a practical choice since the two peptides are small enough that they do not need additional compute resources for training our deep learning model. Similarly, once the training is complete, the same Summit node is also utilized for inference, i.e., to identify novel conformations determined by the CVAE. 

\section{Results}\label{sec:results}
\begin{figure*}[ht!]
  \centering
  \includegraphics[trim=0 0 0 0,clip,width=\textwidth]{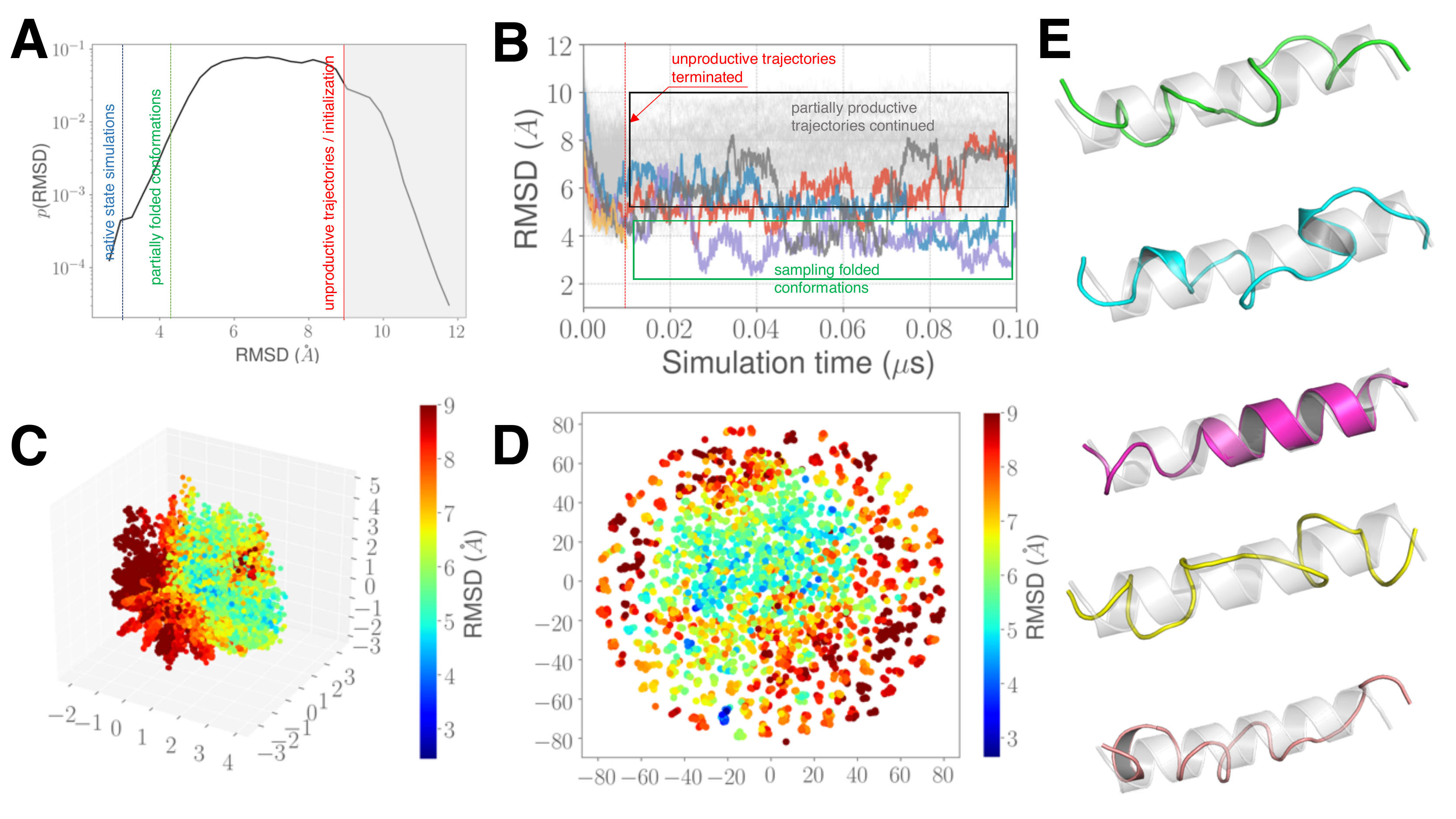}
  \caption{Adaptive simulations of Fs-peptide folding: (A) A summary
  distribution of RMSD (see main text) to the native state of Fs-peptide from
  the adaptive simulations. Note the presence of a large number of
  unproductive trajectories that begin from unfolded state having a RMSD
  higher than 9.2$\AA$. (B) Time evolution of RMSD to the native state from
  the 720 simulations initiated on Summit GPUs using OpenMM simulation engine.
  The productive trajectories (i.e., trajectories that sample conformations
  with $<4.3\AA$ RMSD to the native state at least once) are shown in
  different colors. (C) Summary of the CVAE learned representation shown using
  3 dimensions (for visualization purposes; see main text) showing each
  conformation from the adaptive simulations as a 3D coordinate painted by its
  RMSD to the native state. Notably the states involving the unfolded ensemble
  cluster together, along with intermediate states also clustered together.
  (D) To delineate the folded ensemble, we project the CVAE learned
  presentation onto two dimensions using t-SNE where one can observe the
  separation between the folded and unfolded states. (E) Representative
  conformations from the successful trajectories (in panel B) are shown with
  respect to the native state. The lowest RMSD achieved is about 2.3$\AA$
  shown in magenta. Other conformations sampled from the intermediate states
  are shown for completeness. }
  \label{fig:Fs-peptide-folding}
\end{figure*}

\subsection{Science use cases: Folding simulations of Fs-peptide}

We considered two minimal use cases for our workflow. The first one consisted of folding simulations of the Fs-peptide (21 residues consisting of Ace-A$_5$(AAARA)$_3$A-NME, where Ace and NME represent the N- and C-terminal end caps of the peptide respectively, with A representing the amino acid Alanine and R representing Arginine) in \emph{implicit} solvent conditions driven by a CVAE that learns latent representations from our simulations. Our simulations used the  GBSA-OBC potentials and the AMBER-FF99SB-ILDN force-field set up similar to  previous studies with an aggregate time of 18$\mu$s at 300~K. These simulations were set up in a similar way to previous studies (where each individual simulation was 500 ns); however, the length of any individual simulation in our work was limited to only 50~ns. 

The second set of simulations consisted of a fast folding variant of the villin head piece (VHP; 35 amino acid residues) in \emph{explicit} solvent simulations. These simulations used the AMBER-FF99SB-ILDN with the TIP3P water model, with a cubic box of 60 $\times$ 60 $\times$ 60 $\AA^3$ dimensions. The simulations were carried out for an aggregate of 0.9$\mu$s at 300~K. Note that this timescale is limited by the wall time limits on Summit@OLCF as well as the use of explicit solvent simulations, which can take considerably longer wall clock time to simulate. Individual simulations were capped at 10~ns. In this example we noticed that as a consequence of rather limited sampling, our runs did not end up fully folding VHP to its native state. However, it does allow the simulations to sample partially folded states, where certain $\alpha$-helical turns are formed. As our paper focuses largely in studying the computational performance and scaling aspects of deep learning approaches coupled to MD simulations, we do not present the results from our simulations.  

Fig. \ref{fig:Fs-peptide-folding} summarizes the results of using our workflow in simulating the folding process of Fs-peptide. We first evaluated the quality of folding observed from our simulations using the  root-mean squared deviation (RMSD) with respect to the final folded state (a fully formed $\alpha$-helix) from all of the simulations. As shown in Fig. \ref{fig:Fs-peptide-folding}A, the histogram presents a composite picture of the folding process where by a small proportion of the simulations seem to sample the folded $\alpha$-helical states (labeled native state simulations). Further, a small number of simulations also sample partially folded states (RMSD cut-off of 4.3$\AA$). We also observe that a large portion of the simulations also sample fully unfolded states (RMSD cut-off $> 9.2\AA$). To further understand the time-evolution of the individual trajectories, we plotted the RMSD as a function of simulation time (Fig. \ref{fig:Fs-peptide-folding}B). One of the observations from the aggregate set of simulations is that all of the simulations begin with a high RMSD ($>10\AA$ on average) and evolve gradually towards low RMSD values to the native state. Of the total 720 number of simulations initiated from the unfolded state, 5 of them sample partially folded states where as two of them sample close to the native state of the protein. 

As posited in the beginning of our study, we used our CVAE to drive our enhanced sampling approach. We observed that building a latent representation consisting of 6 dimensions provided the best reconstruction of the simulation data. Since visualization in 6 dimensions is difficult, as shown in Fig. \ref{fig:Fs-peptide-folding}C, we selected 3 dimensions (from the 6) and used it to organize the conformational landscape. Each conformation in the plot is represented as a 3D coordinate and painted with the RMSD to the native state. Note that the RMSD is not part of our training data (only the contact matrices are used as input to train our CVAE) and is an emergent property from our analysis. Most of the unfolded conformations are localized to one region of our representation while many of the folded states are clustered together. We also used t-stochastic neighborhood embedding (t-SNE) to visualize the clustering in a 2D representation. Notably, the folded and unfolded states are separated out (red and blue dots). Additionally, representative structures extracted from the trajectories (from the five trajectories shown in different colors in Fig. \ref{fig:Fs-peptide-folding}B) showed the presence of various intermediates (Fig. \ref{fig:Fs-peptide-folding}E) in the folding process. Note that these states are only a representative subset of the conformations sampled from our adaptive sampling technique.

\subsection{Scaling Profiles for Adaptive Simulations}\label{ssec:folding}

RADICAL Ensemble Toolkit (EnTK) is designed to support the concurrent execution of computational pipelines. Each pipeline is composed of stages and each stage contains an arbitrary set of tasks. Tasks can execute concurrently while stages can execute only sequentially. These properties are insured by design, offering what we have called a Pipeline Stage Task (PST) model for the specification of computational workflows. It is important to note that 'task' here are not functions, methods or sub-processes of one of EnTK components. Task indicates instead a self-contained process (i.e., program) executed and managed by the operating system of the target resource. Consistently, tasks can be a single-threaded, multi-threaded or MPI program, and can use CPUs, GPUs or both within and across the compute nodes of a target machine. 

Specified in PST, the workflow of Fig.~\ref{fig:workflow-meta} consists of a single pipeline with four stages. The first stage executes one or more MD simulations, the second stage aggregates the results, the third stage one or more ML training tasks on the aggregated data produced by the tasks of the first stage, and the fourth stage make an inference about the initial state of the next MD simulation. At this point, the workflow repeats until the protein folds.

\begin{figure}
  \centering
  \includegraphics[trim=0 0 0 0,clip,width=0.49\textwidth]{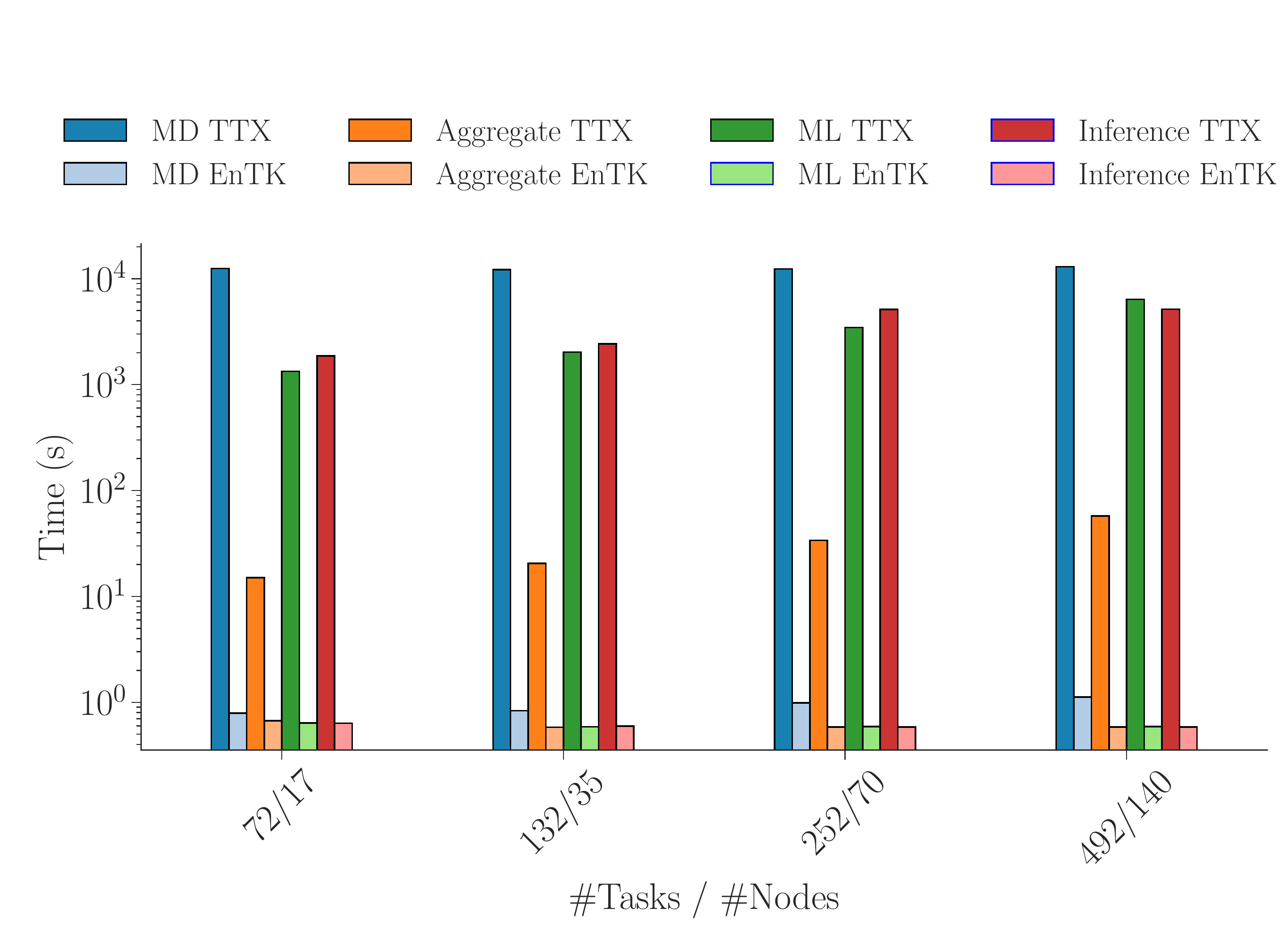}
  \caption{Total time to execution (TTX) and EnTK time overhead of each stage of the workflow described in Fig.\ref{fig:workflow-meta}.}\label{fig:entk}
\end{figure}

EnTK offers two main benefits when implementing this workflow: the ability to arbitrarily change the number of tasks executed in each stage without significant programming or execution time overheads; and the ability to extend the workflow with as many MD simulations/ML inferences stages are required by the protein to fold. Here we focus on the first benefit, studying the relationship between the number of concurrent simulations executed, the amount of data generated, the number of simulated frames and the quality of the leaning we can perform on the produced dataset. In turn, this enables to balance the trade off between resource utilization and the total time required to folding the target protein.

As a first order of concern, we verify that the time overheads of EnTK do not depend on scale and that execution concurrency can also be performed without relevant time overheads. Accordingly, we  designed Experiment 1 to measure how the total execution time (TTX) of each stage of our pipeline changes across number of resources (compute nodes) and number of tasks concurrently executed in each stage. Further, we measured whether and how the EnTK time management overhead (EOH) varies across scales. Note that EnTK performance has been already characterized and that here we aim at confirming a relevant part of the results already published in in Ref.~\cite{balasubramanian2018harnessing}.

Experiment 1 quantifies the relative impact of EnTK on tasks execution and shows how well our runtime system (RADICAL-Pilot) manages execution concurrency. We fix the relation between resources and number of concurrent tasks (i.e., weak scaling), concurrently executing in the first stage 60, 120, 240, 480 and 960 tasks, each using one GPU to execute the OpenMM molecular simulation engine. We do not change the number of tasks on stage 2-4 so to isolate the variations observed by changing the first stage. We execute 1 task in stage 2, 10 tasks in stage 3 and 1 task in stage 4. We execute Experiment 1 on Summit, utilizing between 17 and 280 compute nodes.

Fig.\ref{fig:entk} shows the TTX and EOH for each stage of the pipeline and across the described scales. In absolute terms, TTX of the first stage (MD TTX) weak scales between 60 and 960 tasks/GPUs. The variation of TTX across scales is minimal: 12498s, 12172s, 12378, 12934s, and \ldots for, respectively, 60, 120, 240, 480 and 960 tasks. This indicates that EnTK executed all the tasks concurrently and that the runtime systems added negligible time overheads. EOH is also relatively stable across scales (between 0.79s/\ldots and 60/960 tasks) and, in absolute terms, it is negligible when compared to TTX.

EOH is both stable and negligible across the remaining stages of the pipeline. On the contrary, TTX of Stage 2 (Aggregating TTX) increases with scale due to the number of aggregated files but, comparatively, remains irrelevant compared to the TTX of the other stages. The machine learning tasks of Stage 3 (ML TTX) also increases with scale. This likely depends on the amount of data that needs to be processes in order to train the model. The execution time of Stage 4 also varies with scale but seems to peak at 252/70 tasks. After that, the variation in the execution time is just from 5120s to 5145s.

\begin{figure}
  \centering
  \includegraphics[trim=0 0 0 0,clip,width=0.49\textwidth]{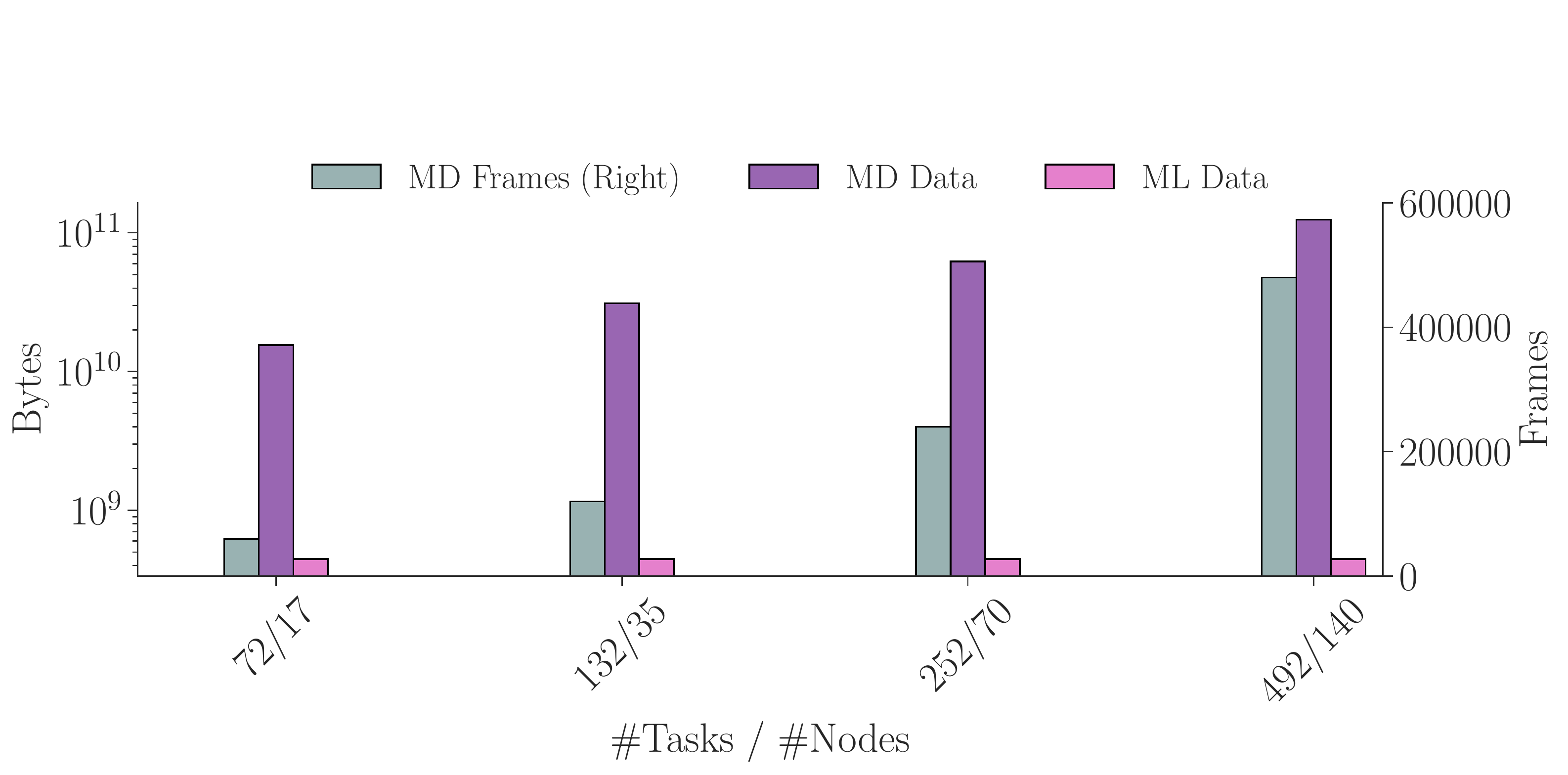}
  \caption{Total amount of data written by each stage of the workflow and total amount of frames calculated by the tasks of Stage0 as a function of the amount of data produced.}\label{fig:data-frames}
\end{figure}

Spending more time to perform the learning and inference tasks is justified only if the quality of the learning and inference processes increases. Increasing the number of MD simulations concurrently executed produces more data and therefore more simulations frames with which to train the ML model. We therefore needed to confirm that the learning and inference processes become more accurate when more data is available. 

\begin{figure}
  \centering
  \includegraphics[trim=0 0 0 0,clip,width=0.49\textwidth]{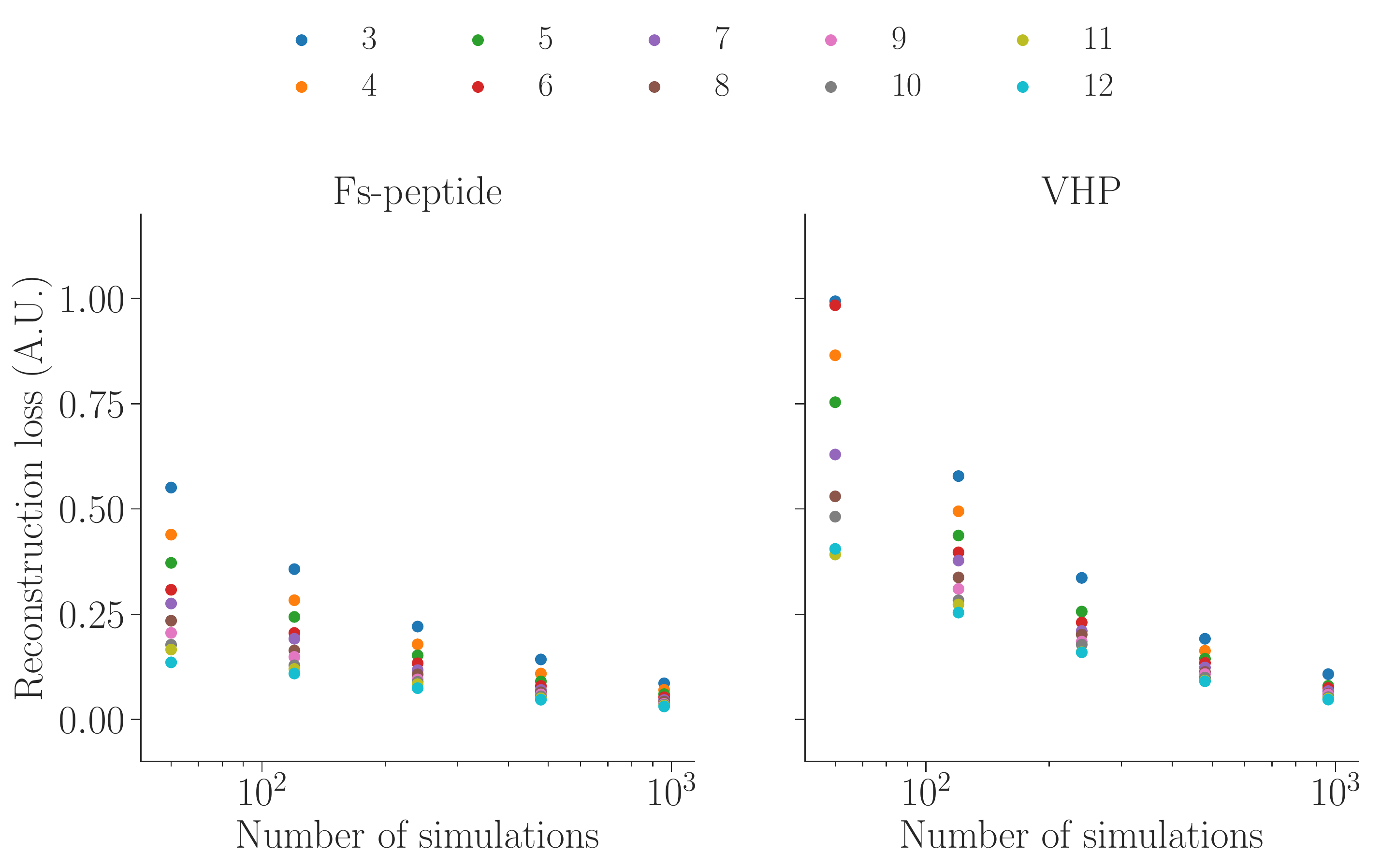}
  \caption{The concomitant increase in the data gathered from the number of concurrent simulations across GPUs improves the quality of learning. The quality of learning is measured using the reconstruction loss for 1,000 frames from the simulations (as a test data that is withheld from training). We show the results for both Fs-peptide and VHP. The results are measured as a function of the number of latent dimensions learned (varying from 3, $\hdots$, 12).}\label{fig:qualityOfLearning}
\end{figure}

Fig.\ref{fig:data-frames} shows the total amount of data produced by the MD
simulations of Stage 1 (purple) and the corresponding amount of frames
calculated (gray, plotted again the right y axis). Data and frames grow with
the number of tasks while the amount of data produced by the learning process
is always the same. To validate if the quality of learning improves with the
concurrent increase in training data, we examined the reconstruction loss
(explained in more detail in ~\cite{bhowmik2018deep}) as a function of both the
latent space dimensions and the number of simulation tasks (on GPUs). Note
that the reconstruction loss measures how well the latent representation is
able to build back the contact matrices after dimensionality reduction using
the CVAE. As shown in Fig. \ref{fig:qualityOfLearning}, as the simulation
tasks increase (along with the total number of conformations), the
reconstruction loss indeed goes down. It is also remarkable that as the number
of dimensions for the latent space representation increases (from 3, $\hdots$,
12), we indeed observe that reconstruction loss decreases -- although for
higher dimensions, this decrease is less pronounced.

This indicates that increasing the training data available for ML approaches can be particularly useful for our proposed adaptive approach. Although we did not present our complete folding process for VHP here, we observed that even with as little as 0.9$\mu$s of aggregate simulations, our adaptive workflow sampled partially folded conformations for VHP. (Note that the folding times for VHP are at the $\mu$s timescales).  


\section{Discussion}
Our efforts in this paper were primarily targeted towards understanding the scaling implications of coupling deep learning approaches with MD simulations. We designed two prototypical workflows involving protein folding simulations, capturing typical use cases of how simple ML approaches such as the CVAE can be coupled with such simulations. Indeed for the case of Fs-peptide under implicit solvent conditions, we could demonstrate that the adaptive sampling approaches can sample folded conformations. On the other hand, the set up of our ML-driven MD workflow could not fully fold VHP in explicit solvent conditions within the time allocated on Summit, but was still able to sample partial folding events in its conformational landscape. 

Our analysis of the Fs-peptide simulations revealed that only 5 out the 720
(including all simulations from steps 2, 3 and 4 of our iterative workflow in
Fig. \ref{fig:workflow-meta}) sampled folded states. A large proportion of
these simulations generated the required training data as part of our
initialization stage (120 simulations, each with 0.1~$\mu$s leading to
12~$\mu$s of sampling). After the training (Stage 2 of the workflow), we were
able to sample the folded states with less than 6~$\mu$s of aggregate
sampling. Without any ML, the aggregate sampling required to fold Fs-peptide
was 14~$\mu$s, which implies that the effective
performance~\cite{fox2019learning} gain in sampling using ML based approaches
is about 2.33 (14$\mu$s to 6$\mu$s). Individual simulations in the ML
driven workflow were only 0.1~$\mu$s in length, as opposed to 0.5~$\mu$s in
traditional (non-ML) sampling, indicating that by culling unproductive
trajectories we can sample the native state of Fs-peptide.


We could have also adopted a previously trained model for our adaptive workflow; however, we explicitly chose not to use such a set up for our experiments. More rigorous tests of the folding times (and kinetics) are required, which we will pursue as part of our future work. This further demonstrates the utility of building ML-driven adaptive MD workflows where some time may be spent initially learning from the running simulations; however, successive iterations of the adaptive workflow can significantly accelerate time-to-solution for expensive simulations. 

To overcome computational costs associated with training the deep learning models (i.e., where the limiting factor is the amount of training data available at the start of the workflow), one may use online machine learning tools~\cite{Ramanathan_2011B}. However, these tools are limited to analyzing limited data streams from MD datasets and may not result in fully transferable models. Therefore, there is also an explicit need to accelerate training for deep learning approaches~\cite{Yoginath_2019} as simulations are concurrently running. In addition, the use of one-shot or few-shot learning approaches~\cite{noe2019boltzmann} can be powerful in overcoming the challenges of having to wait for training iterations to complete. We are planning to pursue these approaches as part of our future work. 

There are several levels at which DL can be interfaced with MD simulations. At
the finest level of granularity, DL models can act as surrogates for
simulations. At the highest level of granularity, reinforcement-based DL
models can serve to steer the computational campaign towards a pre-determined
objective under defined constraints. In between these two different ends of
the spectrum, lies the motif of Fig.\ref{fig:workflow-meta} where DL models
and methods can be used to guide either individual simulations by determining
optimal parameters of exploration, or by intelligently determining regions of
phase space to sample, i.e., enhanced sampling. Needless, to say, these three
levels are not mutually exclusive and can operate concurrently and
collectively to enhanced global computational efficiency, and giving rise to
the concept of Learning Everywhere~\cite{fox2019learning,fox2019understanding}
to enhance computational impact. Although this work investigates and focuses
on the computational motif in Fig.~\ref{fig:workflow-meta}, we will extend
capabilities developed here to cover learning integrated with MD simulations at
all levels.

\bibliographystyle{unsrt}
\bibliography{ReferencesTaxonomyPaper,radical}

\begin{thebibliography}{10}

\bibitem{adcock2006molecular}
Stewart~A Adcock and J~Andrew McCammon.
\newblock Molecular dynamics: survey of methods for simulating the activity of
  proteins.
\newblock {\em Chemical reviews}, 106(5):1589--1615, 2006.

\bibitem{Englander_2014}
S.~Walter Englander and Leland Mayne.
\newblock The nature of protein folding pathways.
\newblock {\em Proceedings of the National Academy of Sciences},
  111(45):15873--15880, 2014.

\bibitem{Shao_2007}
Jianyin Shao, Stephen~W. Tanner, Nephi Thompson, and Thomas~E. Cheatham.
\newblock Clustering molecular dynamics trajectories: 1. characterizing the
  performance of different clustering algorithms.
\newblock {\em Journal of Chemical Theory and Computation}, 3(6):2312--2334, 11
  2007.

\bibitem{Keller_2010}
Bettina Keller, Xavier Daura, and Wilfred~F. van Gunsteren.
\newblock Comparing geometric and kinetic cluster algorithms for molecular
  simulation data.
\newblock {\em The Journal of Chemical Physics}, 132(7):074110, 2010.

\bibitem{Yan_2016}
Yan Li and Zigang Dong.
\newblock Effect of clustering algorithm on establishing markov state model for
  molecular dynamics simulations.
\newblock {\em Journal of Chemical Information and Modeling}, 56(6):1205--1215,
  06 2016.

\bibitem{Shukla_2015}
Diwakar Shukla, Carlos~X. Hern{\'a}ndez, Jeffrey~K. Weber, and Vijay~S. Pande.
\newblock Markov state models provide insights into dynamic modulation of
  protein function.
\newblock {\em Accounts of Chemical Research}, 48(2):414--422, 02 2015.

\bibitem{scherer_pyemma_2015}
Martin~K. Scherer, Benjamin Trendelkamp-Schroer, Fabian Paul, Guillermo
  Pérez-Hernández, Moritz Hoffmann, Nuria Plattner, Christoph Wehmeyer,
  Jan-Hendrik Prinz, and Frank Noé.
\newblock {PyEMMA} 2: {A} {Software} {Package} for {Estimation}, {Validation},
  and {Analysis} of {Markov} {Models}.
\newblock {\em Journal of Chemical Theory and Computation}, 11:5525--5542,
  October 2015.

\bibitem{Ramanathan_2011}
Arvind Ramanathan, Andrej~J. Savol, Christopher~J. Langmead, Pratul~K. Agarwal,
  and Chakra~S. Chennubhotla.
\newblock Discovering conformational sub-states relevant to protein function.
\newblock {\em PLOS ONE}, 6(1):1--16, 01 2011.

\bibitem{bhowmik2018deep}
Debsindhu Bhowmik, Shang Gao, Michael~T Young, and Arvind Ramanathan.
\newblock Deep clustering of protein folding simulations.
\newblock {\em BMC Bioinformatics}, 19(18):484, 2018.

\bibitem{grubmuller1996ligand}
Helmut Grubm{\"u}ller, Berthold Heymann, and Paul Tavan.
\newblock Ligand binding: molecular mechanics calculation of the
  streptavidin-biotin rupture force.
\newblock {\em Science}, 271(5251):997--999, 1996.

\bibitem{abrams2010large}
Cameron~F Abrams and Eric Vanden-Eijnden.
\newblock Large-scale conformational sampling of proteins using
  temperature-accelerated molecular dynamics.
\newblock {\em Proceedings of the National Academy of Sciences},
  107(11):4961--4966, 2010.

\bibitem{kastner2011umbrella}
Johannes K{\"a}stner.
\newblock Umbrella sampling.
\newblock {\em Wiley Interdisciplinary Reviews: Computational Molecular
  Science}, 1(6):932--942, 2011.

\bibitem{reap2018shukla}
Zahra Shamsi, Kevin~J. Cheng, and Diwakar Shukla.
\newblock Reinforcement learning based adaptive sampling: Reaping rewards by
  exploring protein conformational landscapes.
\newblock {\em The Journal of Physical Chemistry B}, 122(35):8386--8395, 2018.

\bibitem{huber1996weighted}
Gary~A Huber and Sangtae Kim.
\newblock Weighted-ensemble brownian dynamics simulations for protein
  association reactions.
\newblock {\em Biophysical journal}, 70(1):97--110, 1996.

\bibitem{Kasson2018-sw}
Peter~M Kasson and Shantenu Jha.
\newblock Adaptive ensemble simulations of biomolecules.
\newblock 13~September 2018.

\bibitem{ma2019deep}
Heng Ma, Debsindhu Bhowmik, Hyungro Lee, Matteo Turilli, Michael~T Young,
  Shantenu Jha, and Arvind Ramanathan.
\newblock Deep generative model driven protein folding simulation.
\newblock {\em arXiv preprint arXiv:1908.00496}, 2019.

\bibitem{galvelis2017neural}
Raimondas Galvelis and Yuji Sugita.
\newblock Neural network and nearest neighbor algorithms for enhancing sampling
  of molecular dynamics.
\newblock {\em Journal of Chemical Theory and Computation}, 13(6):2489--2500,
  2017.

\bibitem{ribeiro2018reweighted}
Jo{\~a}o Marcelo~Lamim Ribeiro, Pablo Bravo, Yihang Wang, and Pratyush Tiwary.
\newblock Reweighted autoencoded variational bayes for enhanced sampling
  (rave).
\newblock {\em The Journal of Chemical Physics}, 149(7):072301, 2018.

\bibitem{Bonati_2019}
Luigi Bonati, Yue-Yu Zhang, and Michele Parrinello.
\newblock Neural networks-based variationally enhanced sampling.
\newblock {\em Proceedings of the National Academy of Sciences},
  116(36):17641--17647, 2019.

\bibitem{noe2019boltzmann}
Frank No{\'e}, Simon Olsson, Jonas K{\"o}hler, and Hao Wu.
\newblock Boltzmann generators: Sampling equilibrium states of many-body
  systems with deep learning.
\newblock {\em Science}, 365(6457):eaaw1147, 2019.

\bibitem{Senior_2018}
R.~Evans, J.~Jumper, J.~Kirkpatrick, L.~Sifre, , T.~F.G. Green, C.~Qin,
  A.~Zidek, A.~Nelson, A.~Bridgland, H.~Penedones, H.~Petersen, K.~Simonyan,
  S.~Crossan, D.T. Jones, D.~Silver, K.~Kavukcuoglu, H.~Hassabis, and A.W.
  Senior.
\newblock De novo structure prediction with deep learning based scoring.
\newblock In {\em Thirteenth Critical Assessment of Techniques for Protein
  Structure Prediction}, 2018.

\bibitem{turilli2019characterizing}
Matteo Turilli, Andre Merzky, Thomas Naughton, Wael Elwasif, and Shantenu Jha.
\newblock Characterizing the performance of executing many-tasks on summit.
\newblock {\em arXiv preprint arXiv:1909.03057}, 2019.

\bibitem{entk-icpp-2016}
Vivek Balasubramanian, Antons Trekalis, Ole Weidner, and Shantenu Jha.
\newblock {Ensemble Toolkit: Scalable and Flexible Execution of Ensembles of
  Tasks}.
\newblock In {\em Proceedings of the 45\textsuperscript{th} International
  Conference on Parallel Processing (ICPP)}, 2016.
\newblock
  \href{http://arxiv.org/abs/1602.00678}{http://arxiv.org/abs/1602.00678}.

\bibitem{balasubramanian2018harnessing}
Vivek Balasubramanian, Matteo Turilli, Weiming Hu, Matthieu Lefebvre, Wenjie
  Lei, Ryan Modrak, Guido Cervone, Jeroen Tromp, and Shantenu Jha.
\newblock Harnessing the power of many: Extensible toolkit for scalable
  ensemble applications.
\newblock In {\em 2018 IEEE International Parallel and Distributed Processing
  Symposium (IPDPS)}, pages 536--545. IEEE, 2018.

\bibitem{merzky2018using}
Andre Merzky, Matteo Turilli, Manuel Maldonado, Mark Santcroos, and Shantenu
  Jha.
\newblock Using pilot systems to execute many task workloads on supercomputers.
\newblock In {\em Workshop on Job Scheduling Strategies for Parallel
  Processing}, pages 61--82. Springer, 2018.

\bibitem{saga-x}
Andre Merzky, Ole Weidner, and Shantenu Jha.
\newblock {SAGA}: A standardized access layer to heterogeneous distributed
  computing infrastructure.
\newblock {\em Software-X}, 2015.
\newblock DOI: 10.1016/j.softx.2015.03.001.

\bibitem{ogf-gfd-90}
Tom Goodale, Shantenu Jha, Hartmut Kaiser, Thilo Kielmann, Pascal Kleijer,
  Andre Merzky, John Shalf, and Christopher Smith.
\newblock {A Simple API for Grid Applications (SAGA)}.
\newblock {OGF Recommendation, GFD.90}, {Open Grid Forum}, 2007.

\bibitem{turilli2017comprehensive}
Matteo Turilli, Mark Santcroos, and Shantenu Jha.
\newblock A comprehensive perspective on pilot-job systems.
\newblock {\em ACM Comput. Surv.}, 51(2):43:1--43:32, April 2018.

\bibitem{turilli2019middleware}
Matteo Turilli, Vivek Balasubramanian, Andre Merzky, Ioannis Paraskevakos, and
  Shantenu Jha.
\newblock Middleware building blocks for workflow systems.
\newblock {\em Computing in Science \& Engineering (CiSE) special issue on
  Incorporating Scientific Workflows in Computing Research Processes}, 2019.

\bibitem{balasubramanian2019radical}
Vivek Balasubramanian, Shantenu Jha, Andr{\'{e}} Merzky, and Matteo Turilli.
\newblock Radical-cybertools: Middleware building blocks for scalable science.
\newblock {\em CoRR}, abs/1904.03085, 2019.

\bibitem{jha2019incorporating}
Shantenu Jha, Scott Lathrop, Jarek Nabrzyski, and Lavanya Ramakrishnan.
\newblock Incorporating scientific workflows in computing research processes.
\newblock {\em Computing in Science and Engineering}, 21(4):4--6, 2019.

\bibitem{fox2019understanding}
Geoffrey Fox and Shantenu Jha.
\newblock Understanding ml driven hpc: Applications and infrastructure.
\newblock {\em arXiv preprint arXiv:1909.02363}, 2019.

\bibitem{fox2019learning}
Geoffrey~C. Fox, James~A. Glazier, J.~C.~S. Kadupitiya, Vikram Jadhao, Minje
  Kim, Judy Qiu, James~P. Sluka, Endre Somogy, Madhav Marathe, Abhijin Adiga,
  Jiangzhuo Chen, Oliver Beckstein, and Shantenu Jha.
\newblock Learning everywhere: Pervasive machine learning for effective
  high-performance computation.
\newblock In {\em {IEEE} International Parallel and Distributed Processing
  Symposium Workshops, {IPDPSW} 2019, Rio de Janeiro, Brazil, May 20-24, 2019},
  pages 422--429, 2019.
\newblock https://arxiv.org/abs/1902.10810.

\bibitem{Ramanathan_2011B}
Arvind Ramanathan, Ji~Oh Yoo, and Christopher~J. Langmead.
\newblock On-the-fly identification of conformational substates from molecular
  dynamics simulations.
\newblock {\em Journal of Chemical Theory and Computation}, 7(3):778--789, 03
  2011.

\bibitem{Yoginath_2019}
Srikanth~B. Yoginath, Maksudul Alam, Arvind Ramanathan, Debsindhu Bhowmik,
  Nouamane Laanait, and Kalyan~S. Perumalla.
\newblock Towards native execution of deep learning on a leadership-class {HPC}
  system.
\newblock In {\em {IEEE} International Parallel and Distributed Processing
  Symposium Workshops, {IPDPSW} 2019, Rio de Janeiro, Brazil, May 20-24, 2019},
  pages 941--950, 2019.

\end{thebibliography}

\end{document}